\documentclass[a4paper]{jpconf}
\usepackage{graphicx}
\usepackage{lineno}
\usepackage{amssymb}
\usepackage{amsmath}
\usepackage[colorlinks]{hyperref}
\begin{document}
\title{Directed flow of $\Lambda$, $\bar{\Lambda}$, $K^{\pm}$, $
  K_S^0$ and $\phi$ mesons 
from Beam Energy Scan Au+Au collisions using the STAR experiment}
\author{Subhash Singha (for the STAR collaboration)}
\address{\it{Department of Physics, Kent State University, Ohio 44242, USA}}
\ead{subhash@rcf.rhic.bnl.gov}
\begin{abstract}
We report the results of $v_1$ and $dv_1/dy$ near mid-rapidity for $\Lambda$, $\bar{\Lambda}$, 
$K^{\pm}$, $K_s^0$ and $\phi$ in Au+Au collisions at $\sqrt{s_{\rm NN}}$ = 7.7,
11.5, 14.5, 19.6, 27 and 39~GeV using the STAR detector at RHIC. The $dv_1/dy$ of $\Lambda$ 
is found to be consistent with that of the proton and shows a change in sign near 
$\sqrt{s_{\rm NN}} = 11.5$~GeV. The $v_{1}$ slope for $\bar{\Lambda}$, $\bar{p}$ and
$\phi$ shows a similar trend for $\sqrt{s_{\rm NN}}$ $>$ 14.5~GeV, while
below 14.5~GeV, $\phi$ $v_{1}$ is consistent with zero but with a large uncertainty.
The $dv_{1}/dy$ for net protons and net kaons is similar for $\sqrt{s_{\rm NN}} >$ 14.5~GeV, 
but they deviate at lower beam energies. 
\end{abstract}
\section{Introduction}
The measurement of collective flow in relativistic heavy-ion
collisions can offer an insight into the Equation of State (EoS) of the
QCD matter produced in early stages of the collision~\cite{earlytime}. 
The directed flow ($v_{1}$) is characterized by the first harmonic
coefficient in the Fourier expansion of the azimuthal distribution of the
produced particles with respect to the reaction plane ($\Psi_R$), i.e, $v_{1}
\equiv \langle \cos(\varphi - \Psi_R) \rangle$, where $\varphi$ denotes the 
azimuthal angle of the produced particles~\cite{flow_method}. Both
hydrodynamic~\cite{hydro_heinz, stocker_npa_750} and transport
model~\cite{urqmd} calculations imply that $v_{1}$ is
sensitive to the EoS and is a promising observable to explore the QCD
phase diagram. One of the main goals of the Beam Energy Scan (BES) program
at RHIC is to search for signatures of a possible QCD critical point and
first-order phase transition~\cite{BES1,BES2,BES-II}. Recently, STAR 
published the beam energy dependence of the slope of directed flow ($dv_1/dy$) 
for protons and net protons near mid-rapidity \cite{star_prl_112}. The 
observation of a minimum in the slope of directed flow for protons and net 
protons around $\sqrt{s_{\rm NN}} =$ 10-20~GeV, and a double sign change in this 
observable for net protons, point to a possible softening of the QCD equation of 
state \cite{jam_attractive}. However, other recent model calculations, with and 
without crossover and first-order phase transitions, show large discrepancies 
with the STAR measurements~\cite{jam_attractive, hybrid_frankfurt,
  3FD, hsd}. 

A new set of measurements with different hadron species ($\Lambda$, $\bar{\Lambda}$,
$K^{\pm}$, $ K_{S}^{0}$ and $\phi$), and hence different constituent
quarks, will not only help to understand the QCD phase transition,
but in addition will allow to disentangle the role of produced and
transported quarks in heavy-ion collisions. In particular, the
$\phi$ meson offers a unique advantage because its mass is similar to
the proton mass, yet it is a vector meson. Moreover $\phi$ meson is
minimally affected by late-stage hadronic interactions.

\section{Analysis Details}
The $v_{1}$ analysis is applied to data taken by the STAR detector during 2010, 
2011 and 2014. The STAR detector offers uniform acceptance, 
full azimuthal coverage, and excellent particle identification. The Time
Projection Chamber (TPC)~\cite{TPC} performed charged particle tracking near
mid-rapidity. The collision centrality is estimated from the charged
particle multiplicity within the pseudorapidity region $|\eta| < $ 0.5. Particles 
are identified using both the Time Projection Chamber (TPC) and Time of Flight 
(TOF)~\cite{TOF} detectors. Two Beam Beam Counters (BBC)~\cite{BBC}, covering 
$3.3 < |\eta| < 5.2$, are used to reconstruct the event plane. The BBC event 
plane is based on the first harmonic, since the $v_{1}$ signal is strong
in the BBC acceptance region at BES energies. Moreover, a large $\eta$-gap 
relative to the TPC reduces non-flow effects \cite{flow_method} in the $v_{1}$ 
measurements. The uncharged particle species are measured through the 
reconstruction of the topology of their weak decays into charged particles: 
$K_s^0 \rightarrow \pi^{+} \pi^{-}$, $\Lambda \rightarrow p \pi$ (and charge 
conjugates for anti-particles). Topological selection cuts are applied in order to
reduce backgrounds without much loss of the
signal. The $\phi$ meson is measured using its hadronic decay channel: $\phi
\rightarrow K^{+}K^{-}$.  The $v_{1}$ for these particles is
extracted by an invariant mass method~\cite{prc_invm}. 
\begin{figure}[ht]
\begin{center}
\includegraphics[width=25pc]{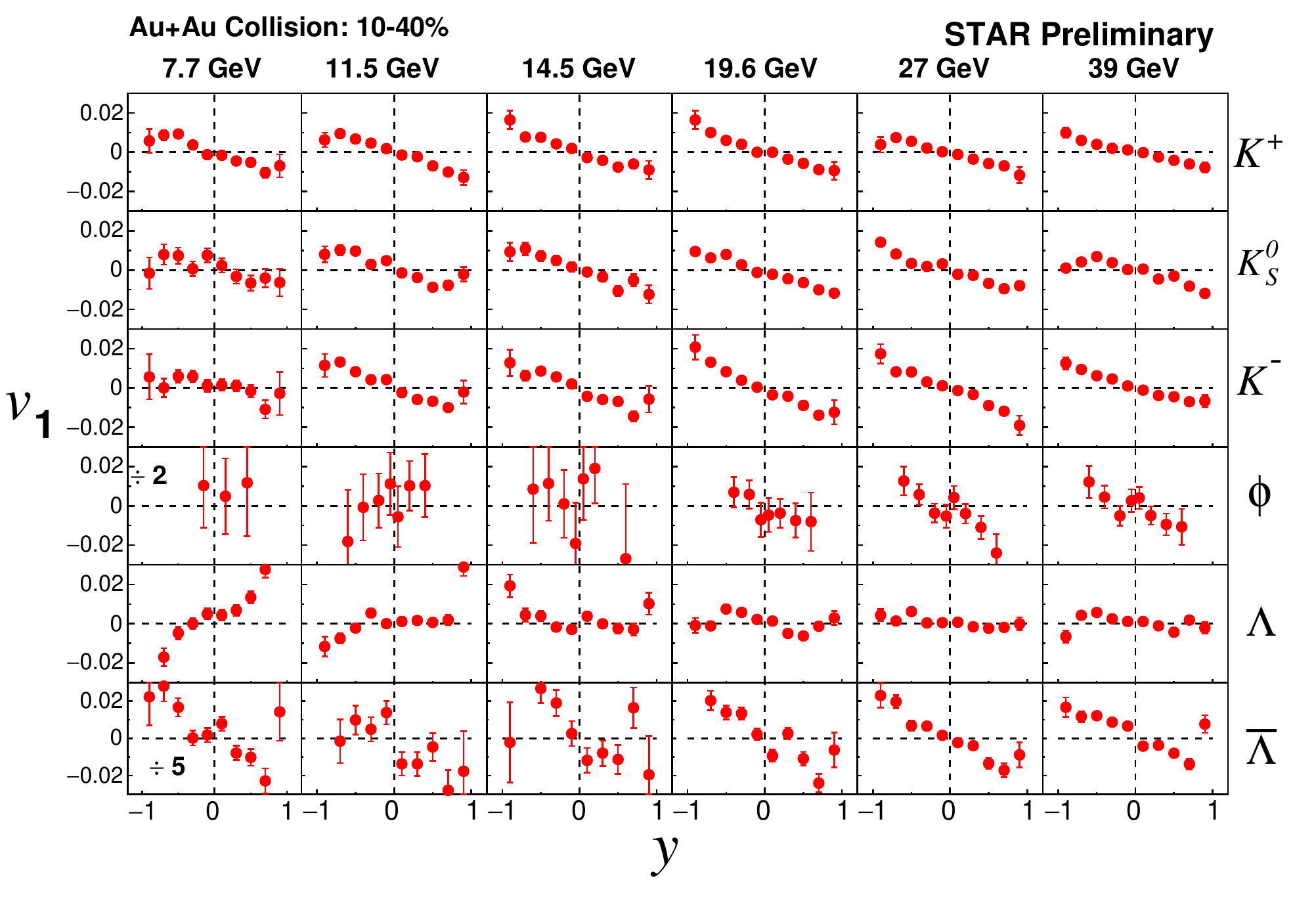}
\caption{\label{v1vsy} Rapidity dependence of $v_{1}$ for $K^{\pm}$,
  $K_s^0$, $\phi$, $\Lambda$ and $\bar{\Lambda}$ in 10-40$\%$ Au+Au collisions at $\sqrt{s_{\rm NN}}$ = 7.7,
11.5, 14.5, 19.6, 27 and 39~GeV.}
\end{center}
\end{figure}
\section{Results and Discussion}
\begin{figure}[ht]
\begin{minipage}{18pc}
\includegraphics[width=18pc]{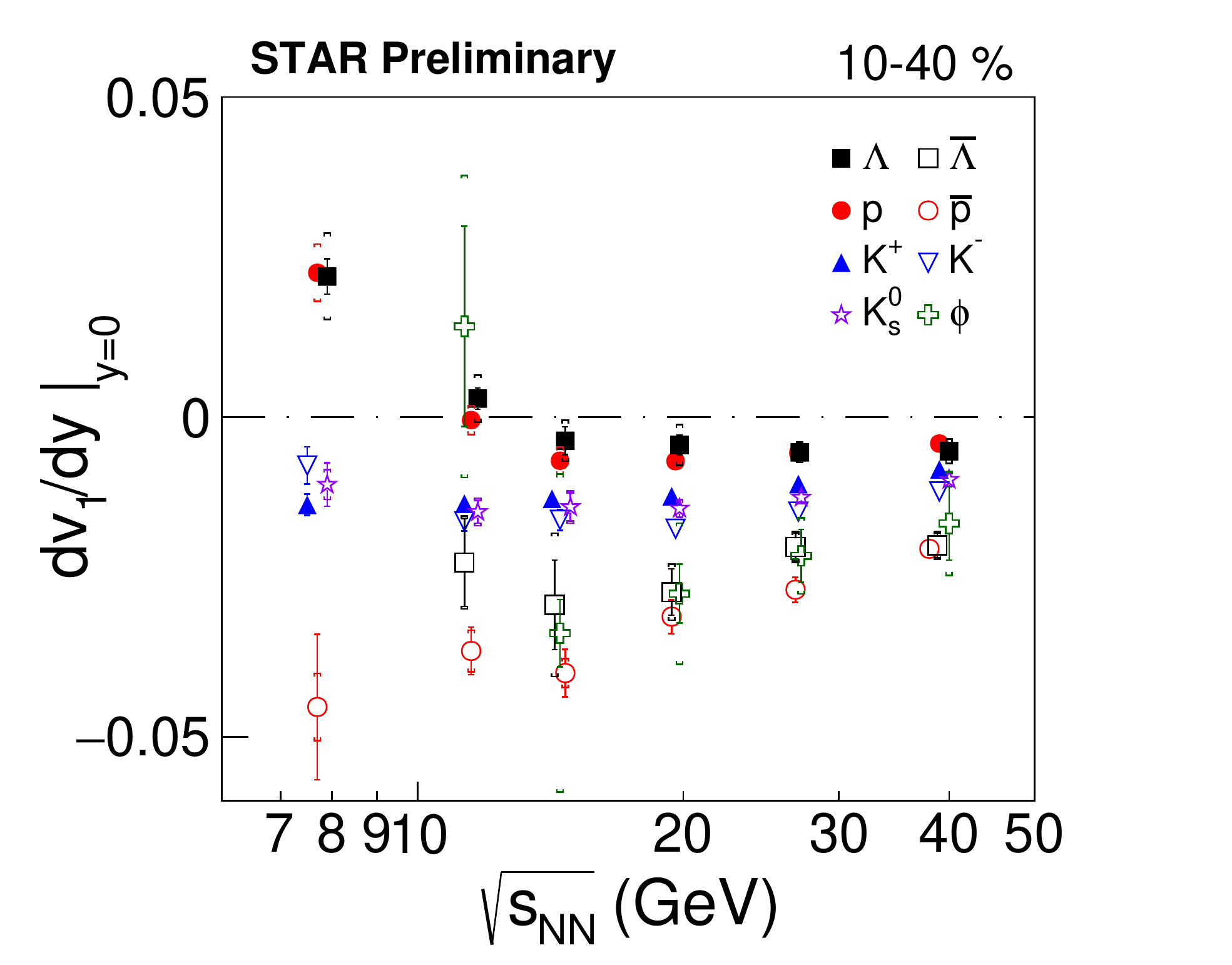}
\caption{\label{dv1dy_alldata} (Color online) Beam energy dependence
  of $dv_{1}/dy$ for $p$, $\bar{p}$, $K^{\pm}$, $K_s^0$, $\phi$, 
  $\Lambda$ and $\bar{\Lambda}$ in 10-40$\%$ Au+Au collisions.}
\end{minipage}\hspace{1pc}
\begin{minipage}{18pc}
\includegraphics[width=18pc]{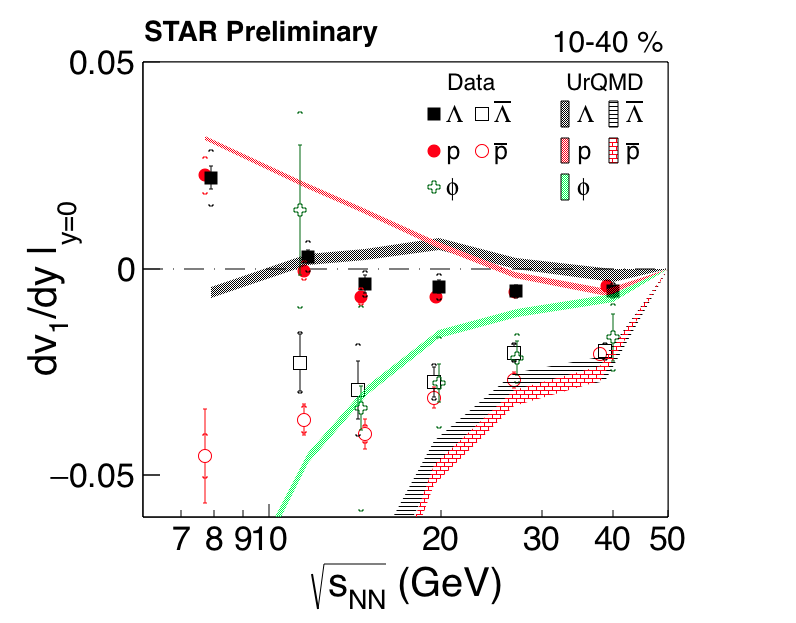}
\caption{\label{dv1dy_urqmd} (Color online) Beam energy dependence of
  $dv_1/dy$  for $p$, $\bar{p}$, $\phi$, $\Lambda$ and $\bar{\Lambda}$
  in 10-40$\%$ Au+Au collisions. STAR data are compared with UrQMD~\cite{urqmd} model calculations.}
\end{minipage}
\end{figure}

\begin{figure}[h]
\begin{minipage}{22pc}
\includegraphics[width=22pc]{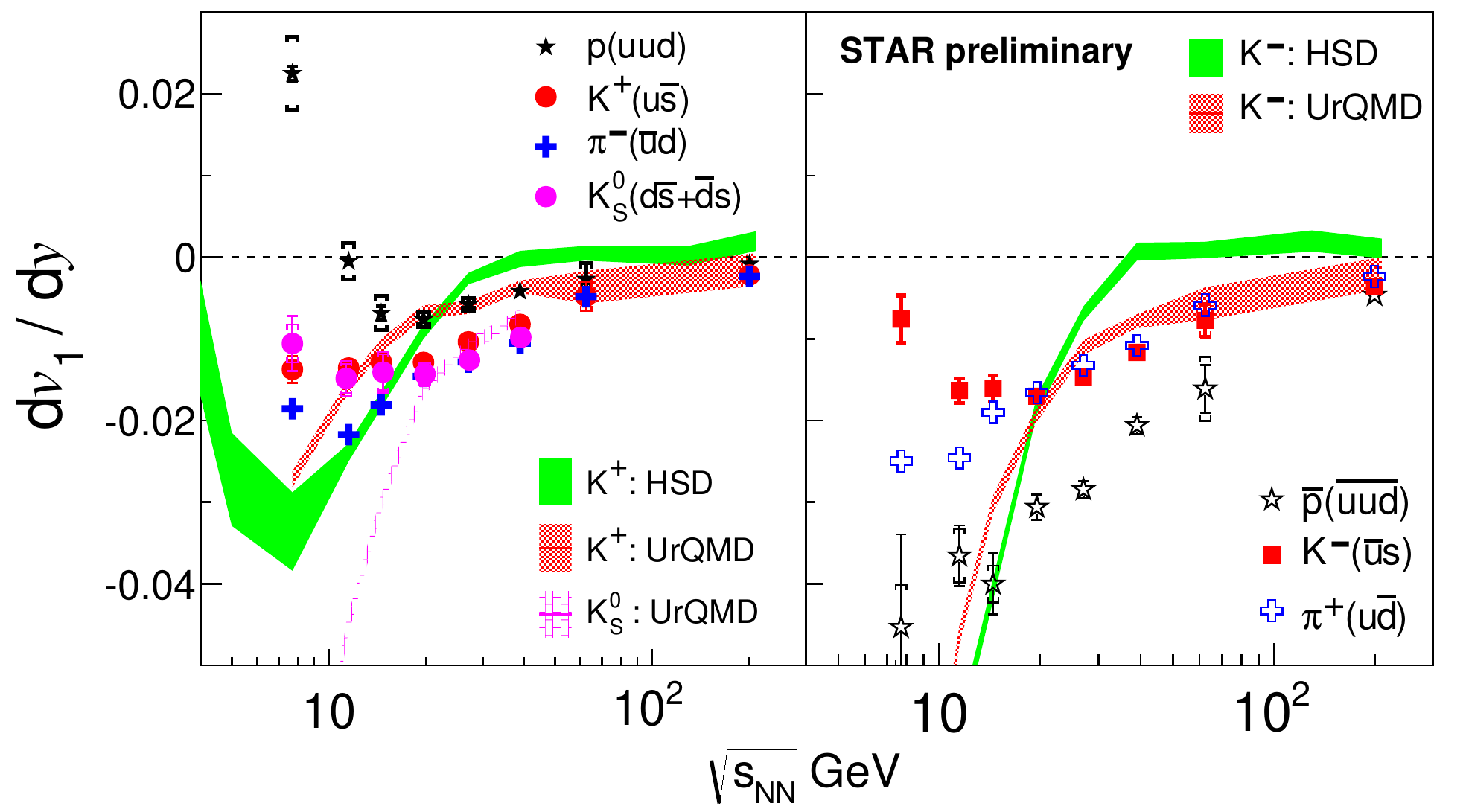}
\caption{\label{dv1dy_2panel} (Color online) Left Panel: Beam energy
  dependence of $dv_1/dy$ for $p$, $K^+$, $\pi^-$ and $K_s^0$ in 10-40\% Au+Au
  collisions. Right panel: Beam energy dependence of $dv_1/dy$ for $\bar{p}$, 
  $K^-$, $\pi^{+}$ in 10-40$\%$ Au+Au collisions. HSD~\cite{hsd} and
  UrQMD~\cite{urqmd} model calculations are shown for $K^\pm$ and $K_s^0$.}
\end{minipage}\hspace{1pc}%
\begin{minipage}{15pc} 
\includegraphics[width=15pc]{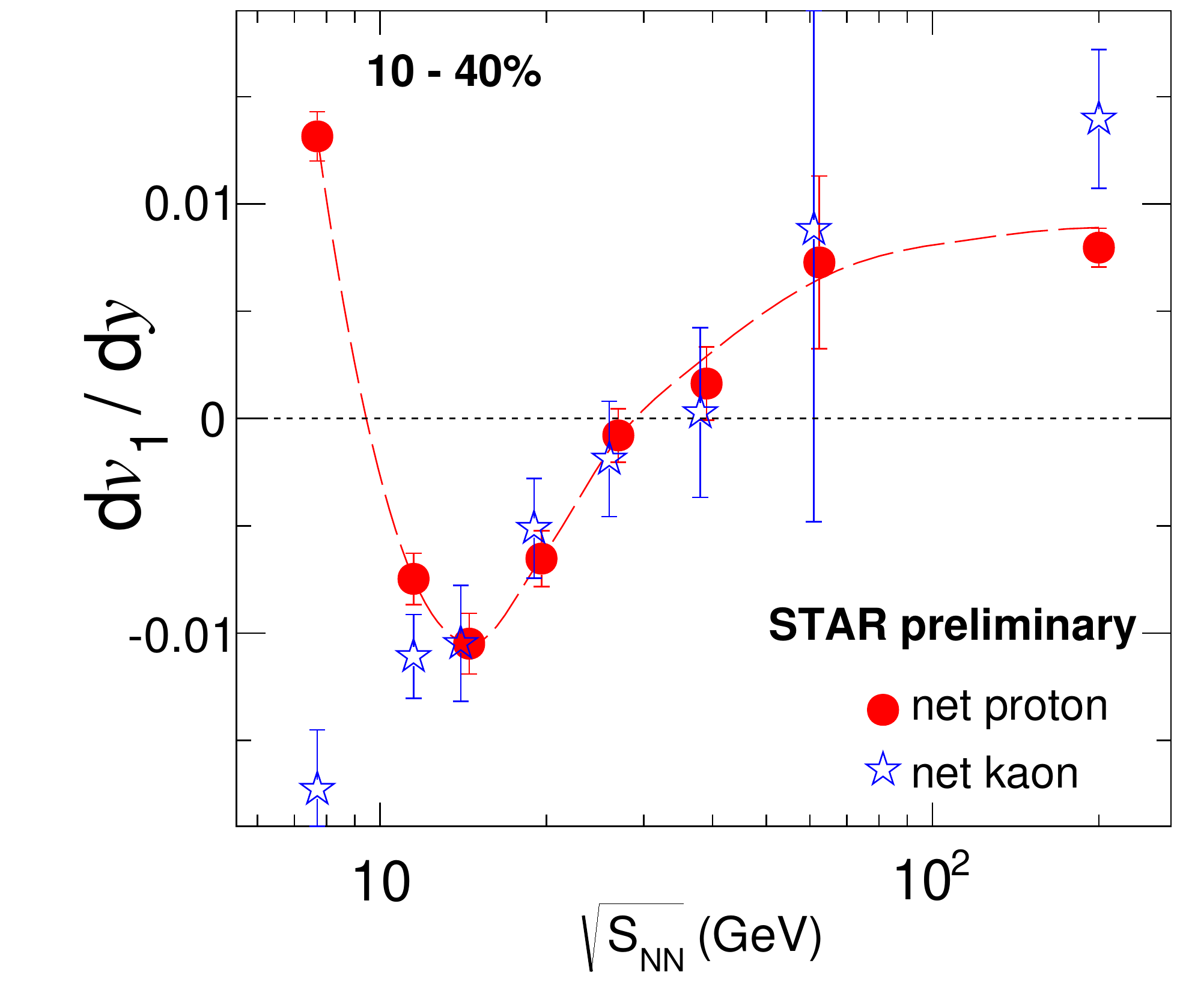}
\caption{\label{netv1} (Color online) Beam Energy dependence of net-particle
  $dv_{1}/dy$ in 10-40$\%$ Au+Au collisions.}
\end{minipage}
\end{figure}

Figure~\ref{v1vsy} presents $v_{1}$ as a function of rapidity for
$K^{\pm}$, $K_s^0$, $\phi$, $\Lambda$ and $\bar{\Lambda}$ in 10-40\%
Au+Au collisions at $\sqrt{s_{\rm NN}}$ = 7.7, 11.5, 14.5, 19.6, 27 and 39
GeV. The data points for $\bar{\Lambda}$ and $\phi$ at 7.7~GeV are
divided by factors of 5 and 2, respectively, in order to use the same 
vertical scale in all panels. In a previous analysis, the $v_1(y)$ slope
parameter near mid-rapidity was extracted by fitting it with
a cubic function \cite{star_prl_112}. Although the cubic fit reduces
sensitivity to the rapidity range where the fit is performed, it
is unstable for particles with poor statistics (e.g. $\phi$ and
$\bar{\Lambda}$). Therefore in Ref.~\cite{star_qm_2015} and the present 
analysis, the $v_1$ slope is based on a linear fit within the range 
$|y|<0.8$ for all particle species and all energies. All kaons and 
$\bar{\Lambda}$ show a negative slope at all beam energies. The $\phi$ 
meson shows a negative slope at $\sqrt{s_{\rm NN}} >$ 14.5~GeV.   
The $dv_{1}/dy$ for all the measured strange hadrons is shown in 
Fig.~\ref{dv1dy_alldata}. 
The $dv_{1}/dy$ of $\Lambda$ is consistent
with that of protons at all beam energies, while the
$\bar{\Lambda}$ follows the trend of antiprotons. The $dv_{1}/dy$
for $K^{+}$ is closer to zero than that for $K^{-}$ for $\sqrt{s_{\rm NN}}
>$ 11.5~GeV, while the trend is reversed at 7.7~GeV. The $K_s^0$ $dv_1/dy$ 
lies between $K^+$ and $K^-$ at all beam energies. Theorists \cite{potentials} 
have argued that kaon-nucleon potentials influence directed flow at these 
energies, where the baryon density is high. The $dv_{1}/dy$ for $\phi$
shows similar trends as $\bar{\Lambda}$ and $\bar{p}$ for $\sqrt{s_{\rm NN}} >$ 
14.5~GeV, while at 11.5~GeV, $dv_{1}/dy$ for $\phi$ is consistent with zero 
with large uncertainty. In the higher energy region where antibaryon production 
is significant, $\bar{p}$, $\phi$ and $\bar{\Lambda}$ show similar energy 
dependence. 
Figure~\ref{dv1dy_urqmd} presents a comparison between the measurements and 
UrQMD~\cite{urqmd} for $p$, $\Lambda$ and $\phi$. UrQMD seems to follow the 
trend of the data for higher beam energies, but deviates at lower energies. 
Figure~\ref{dv1dy_2panel} shows the $dv_{1}/dy$ of $p$, $K^{+}$,
$\pi^{-}$ and $K_S^0$ on the left panel, and corresponding
anti-particles in the right panel, as a function of beam energy. Here the
particles in the left panel are expected to have more quarks from
stopped initial-state nucleons than the antiparticles on the right. The
charged kaons and $K_S^0$ are compared with UrQMD~\cite{urqmd} and HSD~\cite{hsd}
model calculations. Both models qualitatively describe the data at
higher energies, but fail to follow the trend of the data at lower energies.
Figure~\ref{netv1} shows a comparison of the beam energy dependence of 
$dv_{1}/dy$ for net protons and net kaons. Study of net particle $v_1$ is 
motivated by the goal of separating the contributions from produced quarks versus 
quarks that are transported from the initial state. We define 
\begin{equation}
F_p     = r_1(y) F_{\bar{p}} + (1+r_1(y)) F_{{\rm net\mbox{-}}p}  
\end{equation}
\begin{equation}
F_{K^+} = r_2(y) F_{K^-}     + (1+r_2(y)) F_{{\rm net\mbox{-}}K}  
\end{equation}
where $F$ denotes $dv_{1}/dy$ for the indicated particle species; $r_1(y)$ and
$r_2(y)$ are the rapidity dependence of the ratios of the corresponding
antiparticles to particles. It is observed that net-$p$ and net-$K$
$dv_{1}/dy$ follow a similar trend for $\sqrt{s_{\rm NN}}$ = 14.5 - 200
GeV, while they deviate from each other strongly at 7.7~GeV. 

\section{Summary}
In these proceedings, we present $v_1$ and $dv_1/dy$ measurements of strange
hadrons from the STAR experiment at RHIC. The $\Lambda$ $dv_1/dy$ is consistent 
with the proton $dv_1/dy$ and both show a sign change near $\sqrt{s_{\rm NN}} = $ 
11.5~GeV. Charged kaons and $K_s^0$ show negative $v_{1}$ for all studied beam
energies, while the $K_s^0$ lies between the charged
kaons. $\bar{\Lambda}$, $\bar{p}$ and $\phi$ show similar $v_{1}$ slope
for $\sqrt{s_{\rm NN}} > $ 11.5~GeV. The UrQMD and HSD results
qualitatively explain the data at higher beam energies, but fail to
describe the trend observed at lower energies.

\end{document}